\documentclass[fleqn,usenatbib]{mnras}
\usepackage[T1]{fontenc}

\DeclareRobustCommand{\VAN}[3]{#2}
\let\VANthebibliography\thebibliography
\def\thebibliography{\DeclareRobustCommand{\VAN}[3]{##3}\VANthebibliography}

\usepackage{graphicx}
\usepackage{xspace}
\usepackage{todonotes}
\usepackage{mathtools}
\usepackage{amsmath, amssymb}
\usepackage{gensymb}
\usepackage{wasysym}

\title[WIMPs search in clusters]{Annihilating Dark Matter Search with 12 Years of Fermi LAT Data in Nearby Galaxy Clusters}

\author[C. Thorpe-Morgan et al]{Charles Thorpe-Morgan$^1$, Denys Malyshev$^1$, Christoph-Alexander Stegen$^1$, Andrea Santangelo$^1$, \newauthor Josef Jochum$^1$\\
$^{1}$ Institut f{\"u}r Astronomie und Astrophysik T{\"u}bingen, Universit{\"a}t T{\"u}bingen, Sand 1, D-72076 T{\"u}bingen, Germany}

\pubyear{2020}

\begin{document}
\label{firstpage}
\pagerange{\pageref{firstpage}--\pageref{lastpage}}
\maketitle

\begin{abstract}
 Galaxy clusters are the largest virialised objects in the Universe and, as such, have  high dark matter (DM) concentrations. This abundance of dark matter makes them promising targets for indirect DM searches.
 Here we report the details of a search, utilising almost 12~years of Fermi/LAT data, for gamma ray signatures from the pair annihilation of WIMP dark matter
 in the GeV energy band. From this, we present the constraints on the annihilation cross-section for the $b\overline{b}$, $W^+W^-$ and $\gamma\gamma$ channels, derived from the non-detection of a characteristic signal from five nearby, high galactic latitude, galaxy clusters (Centaurus, Coma, Virgo, Perseus and Fornax). We discuss the potential of a boost to the signal due to the presence of substructures in the DM halos of selected objects, as well as the impact of uncertainties in DM profiles on the  presented results.  We assert that the obtained limits are, within a small factor, comparable to the best available limits of those based on Fermi/LAT observations of dwarf spheroidal galaxies.
\end{abstract}


\section{Introduction}
In contemporary astrophysics, dark matter (DM) continues to be one of the greatest unknowns in our understanding of the Universe. The dark matter phenomenon manifests itself seemingly universally, pervading a large range of mass scales and types of object -- from dwarf spheroidal galaxies (dSphs) to, galaxies, clusters of galaxies and features of the early Universe. The latest measurements of Planck suggest that dark matter is incredibly abundant and comprises approximately 26.4\% of the energy density of the Universe~\citep{planck18}. However, despite being the second most abundant component in our universe (after dark energy), very little is known about its physical properties. To attempt to explain this phenomenon, many differing theoretical frameworks exist, with a range of physical motivations and explanations~\citep[see e.g][for a review]{annika12,bauer17,baudis18}. Within one broadly discussed paradigm of these, dark matter is composed of Weakly Interacting Massive Particles (WIMPs) with typical masses in the GeV-TeV range, and is characterised by weak-scale interactions with particles in the Standard Model (SM)~\citep[see e.g.][for a recent review]{Roszkowski_2018}.

WIMPs remain a favoured solution to DM for a number of reasons. Firstly, the relic abundance of DM is naturally obtained through calculations utilising WIMPs with the annihilation cross-section of a weak scale (DM-particles velocity averaged cross-section $\langle\sigma v\rangle_{th}\sim 3\cdot 10^{-26}$~cm$^3$s$^{-1}$) and masses in 10~MeV -- 10~TeV range that have undergone a thermal freeze-out in the early Universe (the so called WIMP miracle~\citet{lw77,feng08}), see \citet{profumo13, Baer_2015} for pedagogical reviews.

Furthermore, WIMP-like particles appear naturally in many Standard Model extensions, including supersymmetric SM extensions~\citep{jungman96,catena14,pddg18} (as e.g. the lightest stable supersymmetric particles), compactified extra-dimensions theories~\citep{cheng02,lorenzana05,kakizaki06,hooper07} (as e.g. the lightest Kaluza-Klein states), GUT-inspired theories~\citep{arcadi16,berlin17,arcadi18} , secluded WIMPs~\citep{pospelov08} and many others~\citep[see e.g.][for reviews]{Roszkowski_2018,arcadi18}.

%%%%%%%%%%%%%%%%%%%%%%%%%%%%%%%
\begin{table*}
\begin{tabular}{|c|c|c|c|c|c|c|c|}
\hline
Cluster                 & \textit{l} & \textit{b} & z  & $r_{s}$ &  $\rho_s$ &Covariance  &Reference\\
                    &[deg] &[deg] & & [kpc]&[$10^5 M_{\astrosun}$/kpc$^3$]&$(\log_{10}r_s,\log_{10}\rho_s)$&\\
\hline
Centaurus & 302.398 & 21.561 & 0.0114 &  470 & 2.13 &$10^{-3}\cdot\begin{pmatrix} 3.7 &-6.3\\-6.3 & 13.65\end{pmatrix}$  &\citet{ettori02} \\ 
Coma & 283.807 & 74.437  & 0.0231 & 360 & 2.75 & $10^{-2}\cdot\begin{pmatrix}31.8 &-47.9\\-47.9&74.0\end{pmatrix}$&\citet{gavazzi09}\\ 
Virgo & 187.697 & 12.337 & 0.0036 &  560 & 0.8 &$10^{-3}\cdot\begin{pmatrix}22.5 &-39.7\\-39.7&71.4\end{pmatrix}$&\citet{mclaughlin99}\\
\hline
Perseus & 150.573 & -13.262 & 0.0179  & 369 &2.73   & -- &\citet{simonescu11}\\ 
Perseus & 150.573 & -13.262 & 0.0179 & 530 & 2.36 &--&\citet{ettori02}\\
\hline
Fornax & 236.712 & -53.640 & 0.0046  & 220 &1.25 &-- &\citet{dw01}(DW01)\\
Fornax & 236.712 & -53.640 & 0.0046 & 98 & 14.5 &--&\cite{rb02}(RB02)\\
Fornax & 236.712 & -53.640 & 0.0046 & 34 &22.0 &--&\cite{sr10}(SR10A10)\\
\hline
\end{tabular}
\caption{Details of the sample of nearby clusters analysed in this work. Coordinates are given in galactic longitude and latitude ( \textit{l} and \textit{b} respectively). The characteristic radii $r_s$ and densities $\rho_s$ of NFW profile (see eq.~\ref{eq:NFW}) are adopted from the corresponding references. Note, that $\rho_s$ corresponds to $\rho_0/4$ for an NFW profile. The covariance matrices show the estimated level of correlation between DM profile parameters, see the text for further details.}
\label{tab:clusters}
\end{table*}
%%%%%%%%%%%%%%%%%%%%%%%%%%%%%%%

Perhaps most importantly though, WIMPs remain viable candidates for both direct and indirect searches, and are thought to decay/annihilate into SM particles with a subsequent production of photons~\citep{cirelli11}. In typical GeV-scale WIMP masses, this photon signal is expected to be of a similar energy to it's progenitor (within the GeV-energy band) which can lead to a detectable excess in the gamma ray flux from DM-dominated astrophysical objects. Generally, this excess is characterised by  non-trivial spatial (determined by the square of the DM-density profile) and spectral (dependent on the type(s) of SM particle(s) it mainly annihilates into- the ``annihilation channels'') profiles.

In this study we focused on the detection of such an excess signal in the GeV band from annihilating WIMP dark matter, utilising Fermi/LAT~\citep{Atwood_2009} data on a number of nearby galaxy clusters. We focused on WIMPs that were predominantly annihilating either through the $b\overline{b}$ or $W^+W^-$ channel, suggested to be dominant for a class of well-motivated Constrained Minimal Supersymmetric SM extensions~\citep{Jeltema_2008}. For the Virgo and Centaurus clusters we also provide  constraints for a direct annihilation of DM into photons  ($\gamma\gamma$), expected to manifest itself via a narrow, line-like spectral feature.

Galaxy clusters (GC) are the largest virialised objects in the Universe, and as such represent attractive laboratories for the study of dark matter. Though located much further away than Dwarf Spheroidal galaxies (dSphs), their very large masses and hence large DM content maintain them as a competitive option for deriving limits on DM particle properties. Similar to dSphs, clusters of galaxies are characterised by low astrophysical background in the GeV band and have relatively well measured properties via strong/weak lensing, X-ray observations or cluster kinematics~\citep[see][and references therein]{bha13}. These well measured properties allow for accurate definition of DM density distributions in each cluster, an essential feature for accurate astrophysical DM study. The location of galaxy clusters however can be problematic for their study. In particular, bright foreground galactic diffuse gamma ray emission can weaken the results and introduce uncertainties connected to the incompleteness  or inaccuracies in the template of this emission. To  minimise the effect of gamma ray galactic diffuse emission, we explicitly selected clusters located at high galactic latitudes ($b>10^\circ$) for the analysis.

Following this introduction the article will be structured as follows. In section~\ref{sec:clusters_sample} we discuss the form of the signal from annihilating DM and its implication on the study. We furthermore outline the approach we took in selecting the clusters for study and highlight those selected. In this section we also discuss the dark matter distribution in each cluster and the implications of uncertainties in this distribution for the expected DM annihilation signal. Section~\ref{sec:data_analysis} is dedicated to the methodology and contains details of the data used and the associated processing of it.  Finally, section~\ref{sec:results_discussion} discusses the obtained limits on the DM annihilation cross-section for all considered channels in the context of previous work, and their potential implications, before our concluding remarks.

\section{Signal and Galaxy Clusters sample}
\label{sec:clusters_sample}
\subsection{Dark Matter Signal}
\label{sec:dm_signal}
Pair annihilation of DM WIMPs ($\chi$) into SM particles ($f$) follows the form: $\chi\bar{\chi}\rightarrow f\bar{f}$ ( where the annihilation channel is named $f\bar{f}$ correspondingly). The subsequent gamma ray radiation spectrum of this annihilation in any DM-dominated object, within the solid angle $d\Omega$, is given by~\citep[see e.g.][]{cirelli11}.

\begin{align}
\label{eq:dm_annih_flux}
&\frac{dF(E)}{d\Omega}\equiv \frac{dN_\gamma}{dEd\Omega} = \frac{dJ/d\Omega}{4\pi\cdot 2m^{2}_{DM}}\times \sum\limits_{f}b_f\cdot \langle\sigma_f v\rangle\cdot\frac{dN^f_{\gamma}}{dE}(E)
\end{align}
Here $m_{DM}$ is the mass of the DM particle $\chi$, $J$ is the object's $J$-factor, a term given by the integral of DM density $\rho$ square over the line of sight:
\begin{equation}
\label{Jfactor}
     dJ/d\Omega = \int\limits_{l.o.s} \rho^{2}(\ell) d\ell
\end{equation}
Of the two parts that make up the product in Eq.~\ref{eq:dm_annih_flux}, the first is proportional to the total number of DM-particle annihilations in the line of sight, while the second term represents the spectrum of one annihilation, averaged over all possible annihilation channels. To achieve this, the sum in the second term is performed over all possible annihilation channels (all possible $f$'s in $\chi\bar{\chi}\rightarrow f\bar{f}$). Moreover, the $b_f$ term (the branching ratio) corresponds to the probability of DM pair annihilation into certain SM particles (termed: the annihilation channel), such that $\sum\limits_f b_f=1$. $\langle\sigma_f v\rangle$ is the dark matter velocity averaged annihilation cross-section $\sigma_f$ for channel $f\bar{f}$. We note that in the subsequent analysis we have assumed a 100\% branching ratio to each annihilation product, thus making the values of $b_f$ and $\langle\sigma_f v\rangle$ redundant. We have however displayed them for posterity, and to show their relevance for studies where multiple annihilation channels are considered.
Finally, $dN_\gamma^{f}/dE$ represents the gamma ray spectrum produced for one annihilation in its corresponding channel.

%%%%%%%%%%%%%%%%%%%%%%%%%%%%%%%%%%%%%%
\begin{figure*}
\includegraphics[width=\textwidth]{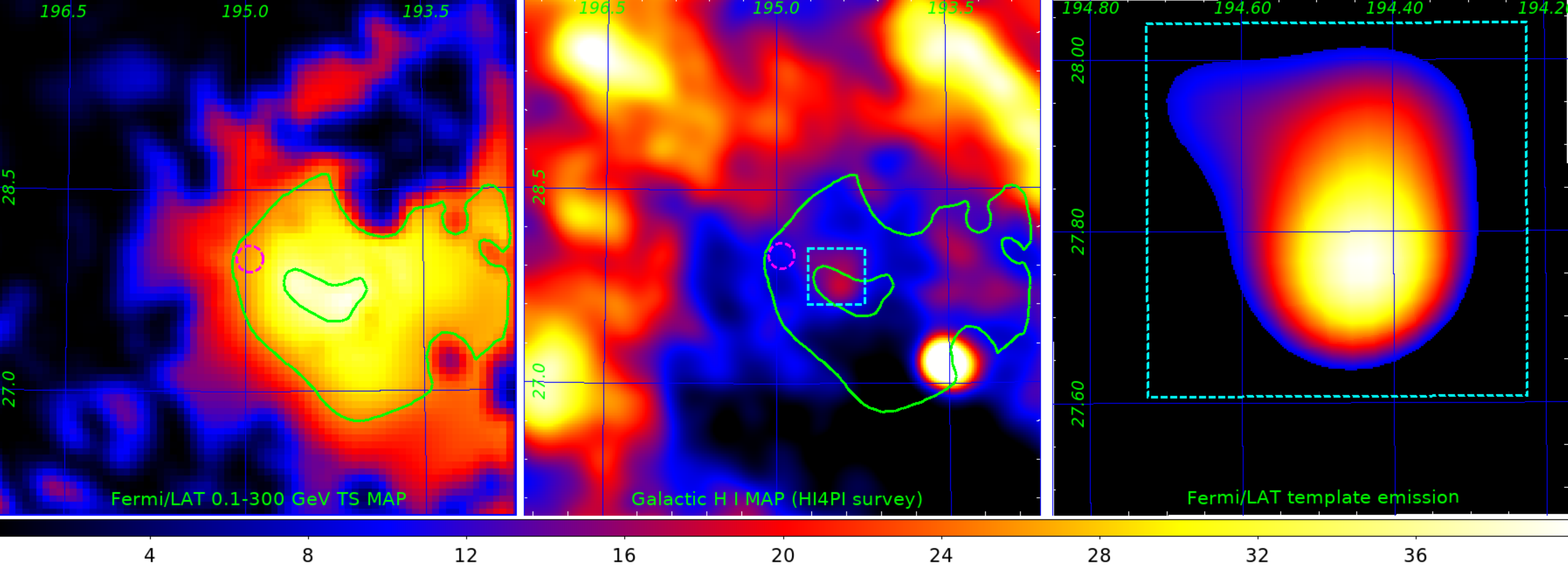}
\caption{Left panel: Fermi/LAT test statistics map of a $2^\circ\times 2^\circ$ region around  the position of the Coma cluster, shown with a dashed magenta circle. The colour scale corresponds to the square of the significance of point-like source added at each point. The bright excess seen is consistent with the results of~\citet{coma_fermi}. Green contours represent isobars of $6\sigma$ and $7\sigma$ detection significance (moving inward respectively). Middle panel: Galactic H~I map of the region from HI4PI survey~\citet{hi4pi}. Green contours are overlayed and correspond to those of the left panel. Right panel: a template added to the model of the Coma cluster region used for Fermi/LAT data analysis. The cyan square corresponds to its counterpart in the middle panel, see text for the details.}
\label{fig:coma}
\end{figure*}
%%%%%%%%%%%%%%%%%%%%%%%%%%%%%%%%%%%%%% 

The expected dark matter annihilation signal from any astrophysical object is thus characterised by a specific spatial shape and spectrum. While the spectrum of the signal is completely determined by ``particle physics'' factors (i.e. by $dN_\gamma^{f}/dE$ terms), its spatial shape is determined by elements of the equation with an astrophysical origin (i.e. dark matter density profile via the $J$-factor term). The overall strength of the signal is determined by the product $J\langle\sigma_f v\rangle$ i.e. is proportional (from Eq.~\ref{Jfactor}) to the square of characteristic DM density in the astrophysical object and DM annihilation cross-section. 
    
This strong, quadratic, dependency  ensures objects with high DM densities have very large $J$-factors and therefore potentially have strong signals from DM annihilation. This makes large objects and those with the aforementioned high DM densities, primary targets for searches of DM-annihilation signals.

\subsection{Clusters' Sample}

As a result of the  previously mentioned criteria for maximising the DM annihilation signal, several types of DM dominated objects are usually presented as favourable targets for searches due to their maximisation of these factors \citep[see e.g.][for a discussion]{bringmann12,funk15}. Among these include nearby dwarf spheroidal galaxies~\citep{archambault17,oakes19,hoof20,linden20}, the Galactic Center~\citep{lucia17,fermi_gc} and clusters of galaxies~\citep{Xiaoyuan,zimmer15,adams16} to name but a few.

In the following, we focus our search for a DM annihilation signal on a sample of nearby galaxy clusters. For our studies we selected a sample of nearby ($z\lesssim 0.02$), high galactic latitude ($|b|>10^\circ$) clusters; for each of these clusters we also utilised dark matter profiles reported in the literature. The sample included 5 objects (The Centaurus, Coma, Virgo, Perseus and Fornax clusters, see Table~\ref{tab:clusters}) and partially intersected with a sample used in a previous study~\citet{Xiaoyuan}. In comparison to this study, our work benefited from substantially better statistics in the data and improved Fermi/LAT calibration, see Sec.~\ref{sec:data_analysis} for details.

\subsection{Dark Matter Distribution}
Given the quadratic dependency of the DM annihilation signal (Eq.~\ref{eq:dm_annih_flux}-\ref{Jfactor}) on the dark matter density distribution, it was imperative for our studies to ensure the dark matter profile was accurate and well described the distribution in the cluster being studied.

For each of the selected clusters we adopted the smooth, spherically symmetric generalized NFW profile (also known as Zhao profile~\citep{zhao96})
\begin{align}
\label{eq:NFW}
& \rho(r) = \frac{2^{\frac{\beta-\gamma}{\alpha}}\rho_s}{(\frac{r}{r_s})^\gamma(1+(\frac{r}{r_s})^{\alpha})^{(\beta-\gamma)/\alpha}} \\ \nonumber
&\mbox{with } \alpha = 1;\quad\beta=3;\quad\gamma=1
\end{align}
which gives the DM density $\rho$ as a function of the radius $r$ from the cluster's center. 
We note that for the quoted $\alpha,\beta,\gamma$ parameters of this profile, Eq.~\ref{eq:NFW} is equivalent to Navarro-Frenk-White~\citep[NFW][]{Navarro_1997} profile with characteristic DM density $\rho_0=4\rho_s$. The characteristic $r_s, \rho_s$ parameters of this profile for the selected sample of clusters are summarised in Table~\ref{tab:clusters} along with the references to the works from which these parameters were obtained.

For NFW profiles~\ref{eq:NFW} the $J$-factor~\ref{Jfactor} is formally divergent at $r\rightarrow 0$ which corresponds to a rapid increase of the NFW DM-profile density at small radii. At radii $r\lesssim 10$~kpc, conversely to this,  observations~\citep{newman13_1, newman13} and simulations~\citep{schaller15,tollet16} of clusters of galaxies both show a flattening of DM density profiles. Such flattening effects can be connected to the effects of baryons/AGN feedback~\citep{gnedin04,teyssier11,castro20,maccio20}. To account for the possible flattening of the DM profile, we explicitly set the DM column density to a constant value within the inner $10$~kpc of each of the considered clusters thus avoiding a nonphysical increase in density at small radii.

The radial dependency of the $J$-factor was derived according to Eq.~\ref{Jfactor} using the CLUMPY v.3 code~\citep{clumpy1,clumpy2,clumpy3}. Utilisation of the CLUMPY code allowed us, in a method consistent over different clusters, to consider the impact of the presence of a large number of substructures in the cluster's DM density distributions, corresponding to the substructures seen in Cold Dark Matter cosmological N-body simulations~\citep[see e.g.][]{diemand07,Springel_2008}. 

The presence of such substructures can significantly boost the expected signal from the outer parts of halos and thus should be properly taken into account. In order to adequately model the substructures, we assumed a mass distribution function of $dN_{sub}/dM\propto M^{-1.9}$ with a 10\% mass fraction in substructures~\citep{Springel_2008}, for the minimal/maximal substructure mass to be $10^{-6}/10^{-2}M_{clust}$  and utilised the same treatment as within \citet{sanchez14} for the subclumps' mass-concentration relation. The substructures' spatial distribution $dN_{sub}/dV$ was selected to follow the host halo's smooth profile.%boosting the total J-factor evenly, rather than lending a greater boost to outer areas like in other subhalo models.

\begin{figure}
\includegraphics[width=0.47\textwidth]{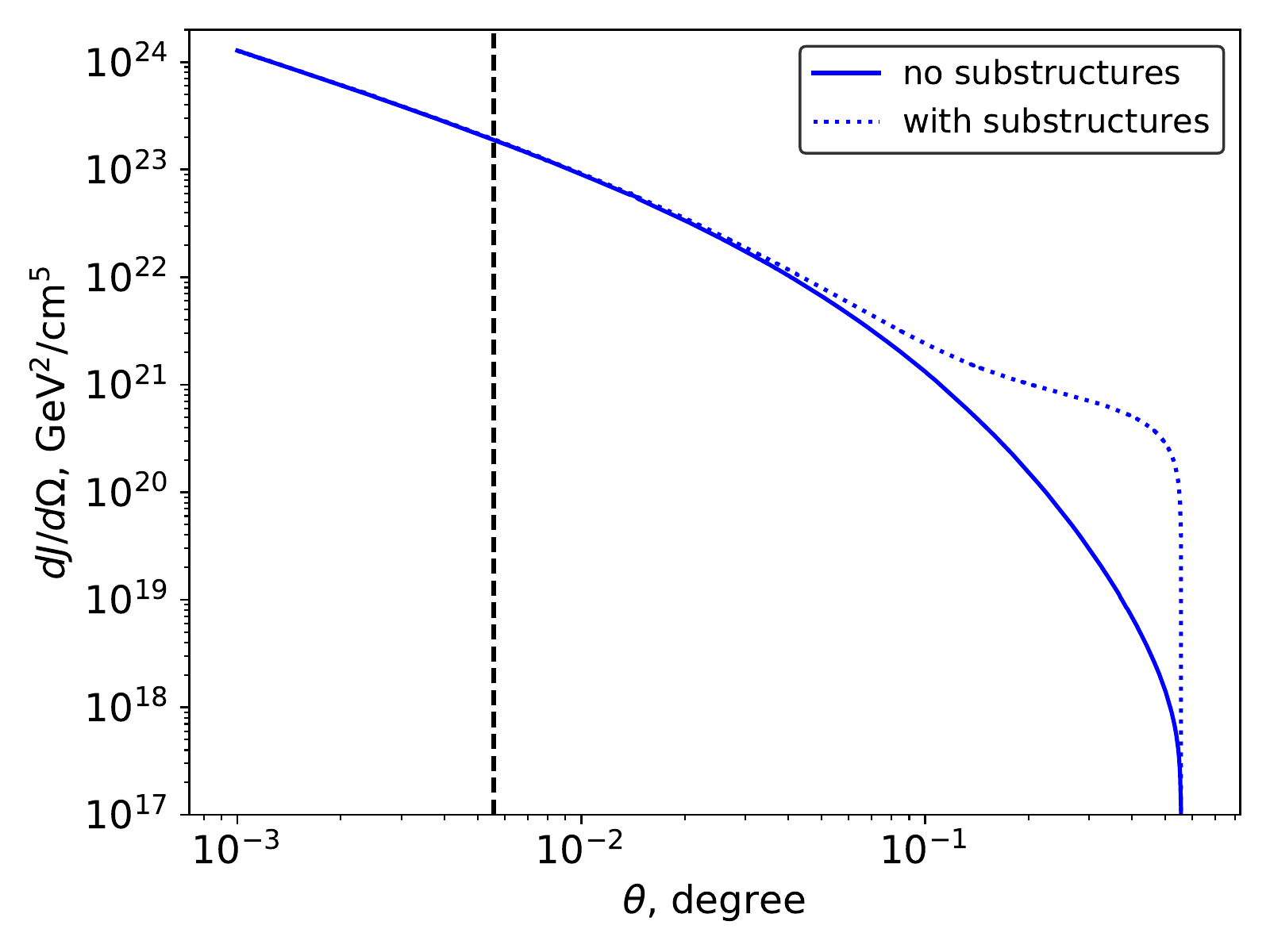}
\caption{ The $J$-factor profile for the Coma cluster is shown here both with and without the presence of substructures, denoted by solid and dotted blue lines in the figure respectively. The vertical dashed line corresponds to a distance of $10$~kpc from the cluster's center; the DM density at radii closer to the center than this point was assumed to be constant. The profile was assumed to continue to  a distance of 1~Mpc from Coma's center, at these greater distances the profile measurements were reported in~\citet{gavazzi09}.}
\label{fig:j_coma}
\end{figure}

Such a choice resulted in a more conservative estimation of the boosted signal (an order of magnitude increase of $J_{tot}\equiv\int (dJ/d\Omega)\, d\Omega$) in comparison to one used by~\citet{Xiaoyuan} (factor of $10^3$). For reference, we show $J$-factors for the Coma cluster with and without the presence of substructures in Fig.~\ref{fig:j_coma}, with solid and dotted blue lines correspondingly. The vertical dashed line corresponds to a distance of $10$~kpc from the cluster's center - the point from which we assumed a constant DM density. 

The density profiles of all clusters were assumed to continue up to the largest distances from the center at which profile measurements were reported in the references in Tab.~\ref{tab:clusters} (0.5~Mpc for Centaurus; 1~Mpc for the rest of the clusters).  We note the possibility of the strong model dependence of the boosted signal thus present the results for both cases below -- the halo with substructures (``boosted'') and smooth halo only (``non-boosted'') dark matter profiles.

%%%%%%%%%%%%%%%%%%%%%%%%%%%%%%%%%%%%%%
%%%%%%%%%%%%%%%%%%%%%%%%%%%%%%%%%%%%%% 

\subsection{Uncertainty Propagation}
\begin{figure*}
\includegraphics[width=0.45\linewidth]{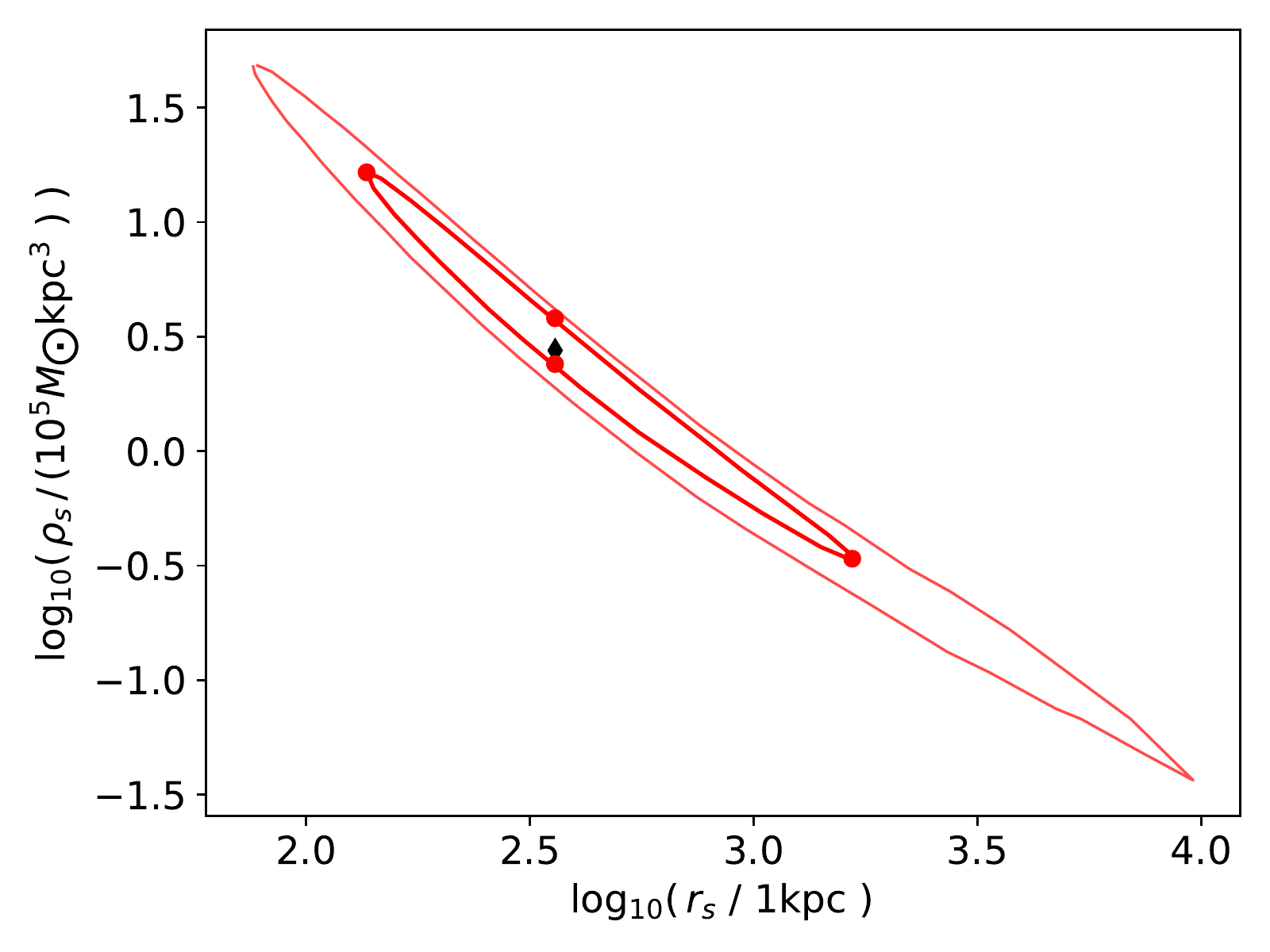}
\includegraphics[width=0.45\linewidth]{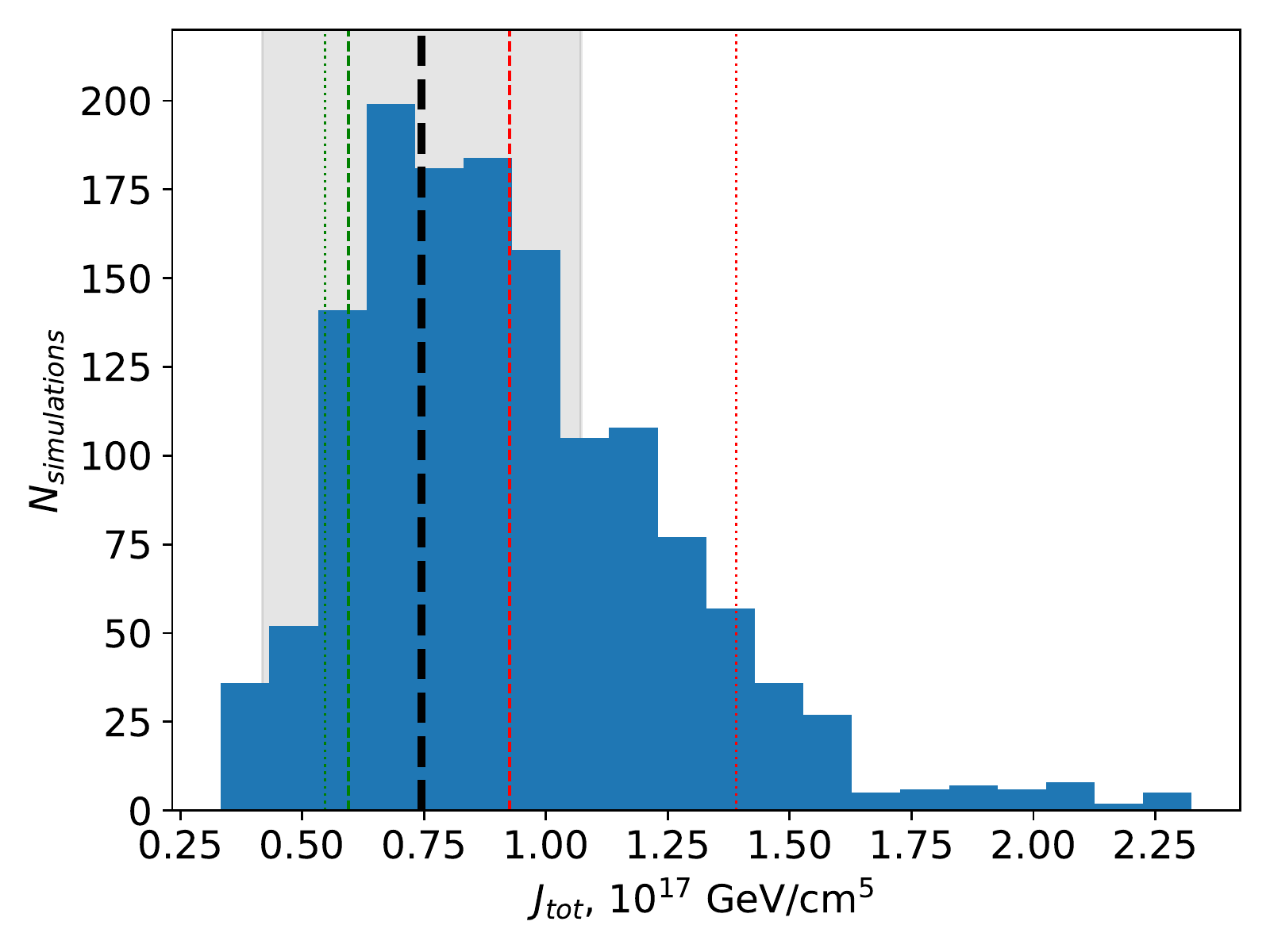}
\caption{Left panel: The red contour represents a $1\sigma$ confidence range for $r_s$, $\rho_s$ parameters of DM profile~\ref{eq:NFW} of the Coma cluster, adopted from the plot within~\citet{gavazzi09}. The central black diamond corresponds to~\citet{gavazzi09} best-fit values of $r_s$ and $\rho_s$. Red square points show the characteristic parameters used for $J$-factor uncertainty estimations, see text for the details.
Right panel: The $J$-factor distribution derived from 1400 points apportioned within the $1\sigma$ and $2\sigma$ contours shown in left panel. The vertical black dashed line shows the mean $J$-factor value corresponding to~\citet{gavazzi09} the best-fit values of $r_s$ and $\rho_s$. The grey shaded region indicates the formal dispersion of the derived distribution plotted around the mean $J$-factor value. Thin dotted lines correspond to red square points in the left panel, see text for the details. }
\label{fig:uncertainties}
\end{figure*}

The total uncertainty in the expected DM annihilation signal~\ref{eq:dm_annih_flux} arises from  uncertainties in the $J$-factor value used for each cluster (see eq.~\ref{Jfactor} ), which is in turn connected to the uncertainties in the cluster's dark matter distribution profile. We note that the $r_s$ and $\rho_s$ profile parameters are usually strongly correlated, see  e.g. Fig.~\ref{fig:uncertainties} (left) showing $1\sigma$ and $2\sigma$ confidence contours (solid and thin red curves) adopted from~\citet{gavazzi09}.
The black diamond symbol here corresponds to the best-fit $(r_s, \rho_s)$ parameters of~\citet{gavazzi09}.

%%%%%%%%%%%%%%%%%%%%%%%%%%%%%%%%%%%%%%
%%%%%%%%%%%%%%%%%%%%%%%%%%%%%%%%%%%%%% 

The strong correlation of the $(r_s, \rho_s)$ parameters reflects the fact that generally, the total masses of clusters are measured far more accurately and frequently ( a value that is roughly proportional to the quantity $\rho_s r_s^{3}$) than the individual values of $r_s$ and $\rho_s$. We note also that the explicit shape of the confidence contours for the DM-profile parameters is not available in the literature for most of the considered clusters. Typically quoted uncertainties for the $r_s$ and $\rho_s$ parameters are given without specification of the confidence contours and normally correspond to the projections of the confidence contour on to the $r_s$ and $\rho_s$ axes, see Fig.~\ref{fig:uncertainties} (left). Although these projections reflect the highest possible uncertainties for the $r_s$ and $\rho_s$ parameters, the direct propagation of these uncertainties on the $J$-factor, without accounting for the correlation of these parameters, can result in drastically overestimating the $J$-factor uncertainty. 

Thus, the usual way of estimating $J$-factor uncertainty $dJ$ is connected to the propagation of uncertainty from the total cluster mass estimation, see e.g.~\cite{Xiaoyuan}. This is equivalent to the propagation of only the density uncertainty for the fixed (to its best-fit value) $r_s$. In terms of Fig.~\ref{fig:uncertainties} the $J-dJ$ and $J+dJ$ are achieved at red points below and above best-fit black diamond point.

In Fig.~\ref{fig:uncertainties} (right) we show a $J$-factor ($J_{tot}\equiv\int(dJ/d\Omega)\,d\Omega$) distribution from simulations based on confidence regions adopted from~\citet{gavazzi09} for the Coma cluster. For the simulations we considered 1000 ($r_s, \rho_s$) (log)uniformly distributed points within the  $1\sigma$ contour and, 400 points (log)uniformly distributed between the $1\sigma$ and $2\sigma$ contours. The selected uniform distribution was chosen in absence of detailed information regarding the probability distribution within the aforementioned contours. For each considered pair ($r_s, \rho_s$) we calculated a $J$-factor in a method identical to that described above for the mean $J$-factor value. The obtained distribution is shown with blue histogram with a solid black dashed line illustrating the mean $J$-factor value (corresponding to the black diamond point in left panel of this figure). The values of $J$-factors corresponding to the estimations based on ``total mass uncertainty'' propagation are shown with dotted green and red lines. The light-gray shaded region depicts the dispersion of the $J$-factors distribution, plotted around the mean $J$-factor value.

While estimation based on the total mass uncertainty provides a reasonable estimation of $J$-factor uncertainty (comparable to the estimation based on the dispersion of the distribution), we note that the corresponding $J$-factors confidence range is somewhat biased to higher $J$-factors. This can consequently result in biased estimations of the uncertainties on the limits of dark matter's annihilation cross-section. Therefore, in what follows, we utilise the uncertainties on $J$-factors based on the dispersion of the $J$-factors distributions rather than ``total mass estimations''.

In addition to uncertainty connected to the $J$-factor value we consider uncertainties in the dark matter annihilation cross-section, connected to an imperfect knowledge of the spatial shape of the expected signal. To estimate the level of this uncertainty we utilise templates for the signal template based on \textit{(i)} best-fit values of $r_s$ and $\rho_s$ presented in the literature (see Table~\ref{tab:clusters}) ; \textit{(ii)} templates based on $(r_s ;\rho_s)$ parameters at the edges of $1\sigma$ confidence region (i.e.the left and right-most red points on the red contour in the left panel of Fig.~\ref{fig:uncertainties}). We argue that these points correspond to the substantially different \textit{spatial} shapes in the DM distribution and should be considered for the estimation of $J$-factor uncertainty connected to the possible spatial variations of the signal. The $J$-factors corresponding to these points are illustrated in the right panel with thin green and red dashed lines.

To be conservative, we used a maximum of two of the uncertainties described above to estimate the uncertainties for the presented limits on the dark matter annihilation cross-section, for each of the considered DM masses.

Among all the clusters listed in Table~\ref{tab:clusters} the explicit information on the correlation of the $(r_s, \rho_s)$ parameters is available only for the Coma cluster. To perform estimations for the Centaurus and Virgo clusters we assume the $1\sigma$ confidence contours to be ellipses on the $\log r_s$ -- $\log \rho_s$ plane (similar to the Coma cluster). The major axes were estimated from the uncertainties of the $r_s$ and $\rho_s$ parameters (i.e. projections of major axes on $\log r_s$, $\log \rho_s$ axes) quoted in the literature. The minor axes and inclination angles of these ellipses were estimated from the total cluster mass uncertainty range presented in literature. 

The aforementioned ellipse-like confidence regions can be described in terms of covariance matrices for $\log r_s$ -- $\log \rho_s$ parameters, see Table~\ref{tab:clusters}. The major and minor axes of each $1\sigma$ confidence ellipse is given by the square root of the eigenvalue of these matrices. The inclinations of ellipses are given by the angle between the eigenvector corresponding to the largest eigenvalue, and $r_s$ axis. For completeness in Table~\ref{tab:clusters} we present also the covariance matrix for the Coma cluster, corresponding to an ellipse approximately fitting the $1\sigma$ red contour from Fig.~\ref{fig:uncertainties}, left panel.

For the Perseus and Fornax clusters we noted several profiles independently derived in the literature, see Tab.~\ref{tab:clusters}. For these objects we estimated uncertainties on dark matter annihilation cross-section limit based on the corresponding quoted best-fit profiles. Such an approach allows one to access the level of systematic uncertainty, known to be dominant at least for the Fornax cluster, see e.g.~\citet{Abramowski_2012}.

In addition to the sources of systematic error discussed above, we would like to note several other possible sources of such errors. These include: the inner core radius of DM density profiles; minimal/maximal mass of the DM halo substructures; mass and spatial distribution of these objects and the mass-concentration model for subhalos. Although the assumptions made in this study regarding the properties of these parameters are rather conservative, we further investigated the effect on the derived J-factors of parameter variations within a factor of 2 (typical for e.g. DM profile parameters). Variations of inner core radius result in a $\sim 10$\% variation of the total J-factor $J_{tot}$ and we find  similar uncertainties applicable to all results presented below. We note however, that this systematic uncertainty is subordinate in comparison to the DM profiles uncertainties discussed above.

In the presence of substructures, the variations of minimal/maximal mass of subhalos result in $\sim 5$\% variations of $J_{tot}$ (applicable to ``boosted'' limits below). The largest uncertainties on $J_{tot}$, reaching a factor of 100, arise from the uncertainties on the mass/spatial distribution and mass-concentration model of subhalos, see~\citet{pinzke09, fermi10, S_nchez_Conde_2011, Pinzke_2011}. We stress, that in what follows we use a conservative estimations for $J_{tot}$ boost due to the presence of subhalos. Utilisation of optimistic parameters can, however, substantially improve derived limits.

\section{Data Analysis}
\label{sec:data_analysis}
This study utilised approximately 12 years worth of Fermi/LAT survey mode data taken from the 4th August 2008 to the 23rd April 2020. The presented analysis was performed with \texttt{fermitools} v.1.2.1 for \texttt{P8R3\_CLEAN\_V2} gamma ray events \footnote{See \href{https://fermi.gsfc.nasa.gov/ssc/data/analysis/documentation/Cicerone/Cicerone_Data/LAT_DP.html}{Fermi/LAT data analysis guidelines.}} within energy range of 100 MeV to 300 GeV.

We applied the standard time cuts as described in \cite{Atwood_2009} as well as a zenith angle cut at $\theta < 100\degree$ to avoid contamination of the data from the Earth's albedo. The binned analysis (with enabled energy dispersion handling) was performed for the events within $15^\circ$ around the position of each of the considered clusters. The model of the region included sources from the 4FGL catalogue~\citep{4fgl_Cat} as well as templates for galactic and extragalactic diffuse emissions given by the \texttt{gll\_iem\_v07.fits} and \texttt{iso\_P8R3\_CLEAN\_V2\_v1.txt} templates correspondingly. The spectral parameters of these sources were initially assumed to be free. In addition we included sources from the 4FGL catalogue up to $10^\circ$ beyond the considered region of interest into the model, with all their parameters frozen to their catalogue values, in order to reduce bias connected to possible presence of bright sources outside of the considered region and effects to do with the LAT's poor PSF at low $\sim 0.1$~GeV energies. 

The spectral parameters of all free sources were determined from the broad band fit of all the available data to the described model. At the second stage of our analysis we fixed the spectral parameters of all sources, except normalisations, to their best-fit values and added  a template for a putative dark matter annihilation emission to the model.

This emission was modelled as a diffuse source with spatial emission distribution proportional to the $J$-factor described in previous sections. The spectral component of the signal was given by the spectrum of annihilating dark matter for the considered ($b\overline{b}$ and $W^+W^-$) channel, provided within fermitools as \texttt{DMFitFunction} based on \citet{Jeltema_2008}. In the case of the $\gamma\gamma$ annihilation channel, we explicitly assumed the spectral part of the signal to be a narrow (5\% energy width) Gaussian line. To account for such a narrow feature we performed the binned analysis with 50 log-equal energy bins per decade of energy (vs. 10 bins per decade for other annihilation channels). For this case, we noted a significant increase in the required computational resources. Thus, we performed the search for the $\gamma\gamma$ annihilation signal only in the Virgo and Centaurus clusters given their superior constraints in other annihilation channels.

\noindent\textit{A note on the Coma cluster.}\\
\label{sec:coma}
Analysing the data on the Coma cluster we note the presence of a weak signal close to the location of the cluster. The presence of a similar signal was reported in~\citet{coma_fermi}. In Fig.~\ref{fig:coma} (left panel) we show the test-statistics (which corresponds to the square of the significance of an added point-like source) map of a $2^\circ\times2^\circ$ region ($0.05^\circ$ pixel size) around the position of the Coma cluster (shown with dashed magenta circle) in the energy range of 0.1--300~GeV. This map is consistent with the results of~\citet{coma_fermi}. The bright residuals represent an emission of a source(s) not included in the 4FGL catalogue. While generally the origin of this emission is not clear~\citep{coma_fermi,liang18} we note the presence of spatially coincident excess in the galactic hydrogen H~I survey HI4PI~\citep{hi4pi}, see middle panel of Fig.~\ref{fig:coma}. Note also, that all intensity variations in the middle panel are within 10-15\% of its mean value. Green contours in the left and middle panels correspond to the $6\sigma$ and $7\sigma$ levels of point-like source detection significance in the Fermi/LAT data. 

Although detailed studies of the observed excess are beyond the scope of this paper, we argue on its potential association with a foreground galactic H~I cloud. In our studies we modelled the excess with a spatial template based on H~I map (see right panel of Fig.~\ref{fig:coma}; dashed cyan squares show identical regions on the right and middle panels). The spectrum of this source was assumed to be a powerlaw (best-fit index $2.65\pm 0.1$, consistent with~\citet{liang18}). We note also that the obtained, relatively soft, best-fit slope is consistent with the reported values for the average spectral slope of diffuse emission at corresponding galactic latitudes, and in several nearby molecular clouds~\citep[see e.g.][]{neronov15,yang16,neronov17,felix20}. Such similarity marginally supports the proposed possible association of the observed emission.

%%%%%%%%%%%%%%%%%%%%%%%%%%%%%%%%%%%%%%
\begin{figure*}
\includegraphics[width=0.45\linewidth]{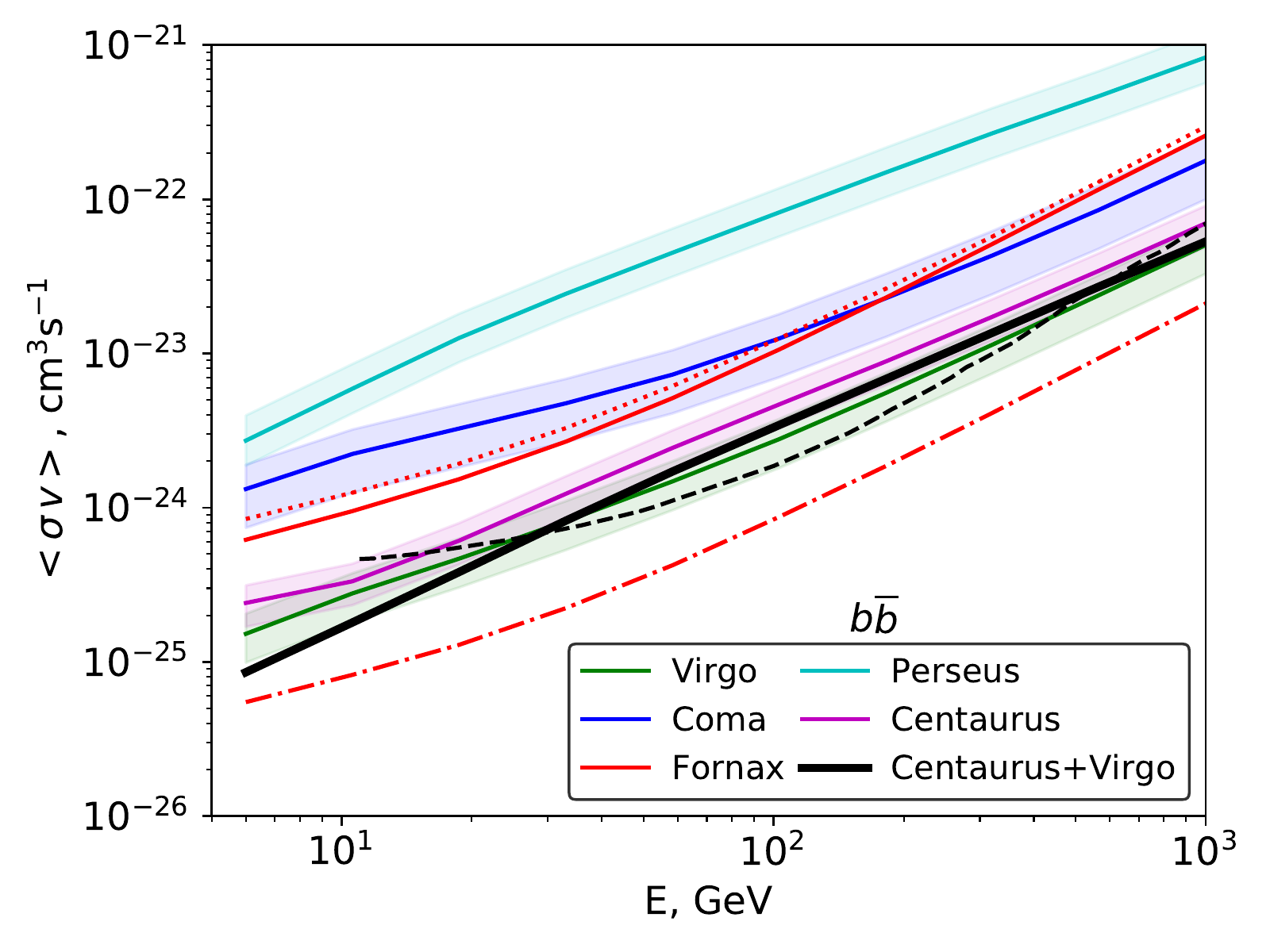}
\includegraphics[width=0.45\linewidth]{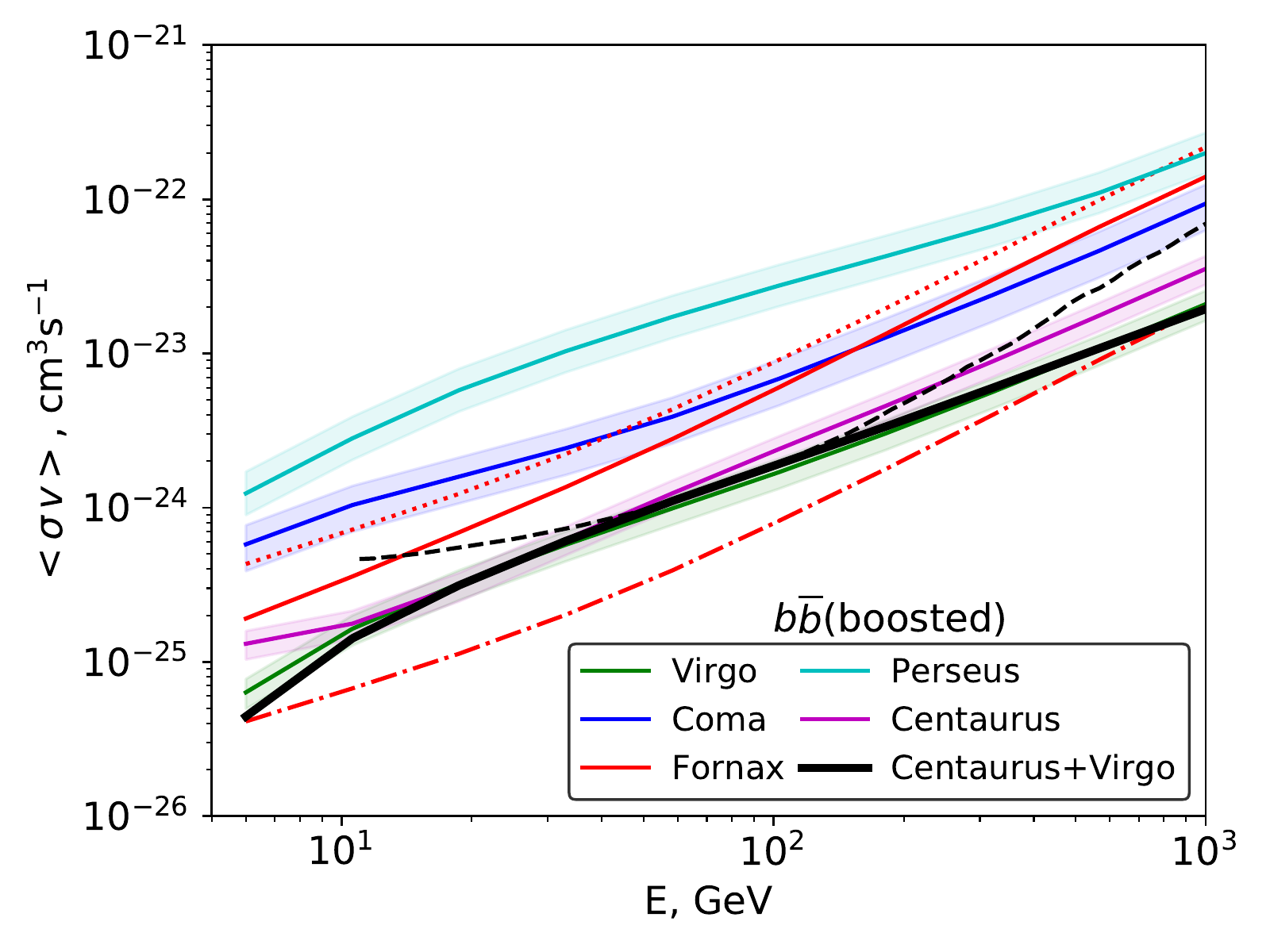}
\caption{The 95\% C.L. limits on the annihilation cross-section of WIMP dark matter for the $b \overline{b}$ annihilation channel, with results shown for the sample of considered galaxy clusters.
The left panel shows the results for smooth DM halos, while the right one shows the results when the presence of substructures is included.
Shaded regions in both panels represent the expected $1\sigma$ uncertainties connected to uncertainties in DM profiles. For the Fornax cluster profiles from literature result in substantially different $J$-factors and consequently different limits on the annihilation cross-section. Results for DW01, RB02 and SR10A10 profiles (see Table~\ref{tab:clusters}) are shown with solid, dot-dashed and dotted red lines respectively. The solid black line corresponds to the limits from a stacked analysis of the Virgo and Centaurus clusters. The thin black dashed line in the left panel represent results of a previous study by~\citet{Xiaoyuan}). }
\label{fig:limits_bb}
\end{figure*}
%%%%%%%%%%%%%%%%%%%%%%%%%%%%%%%%%%%%%% 
 
 %%%%%%%%%%%%%%%%%%%%%%%%%%%%%%%%%%%%%%
\begin{figure*}
\includegraphics[width=0.45\linewidth]{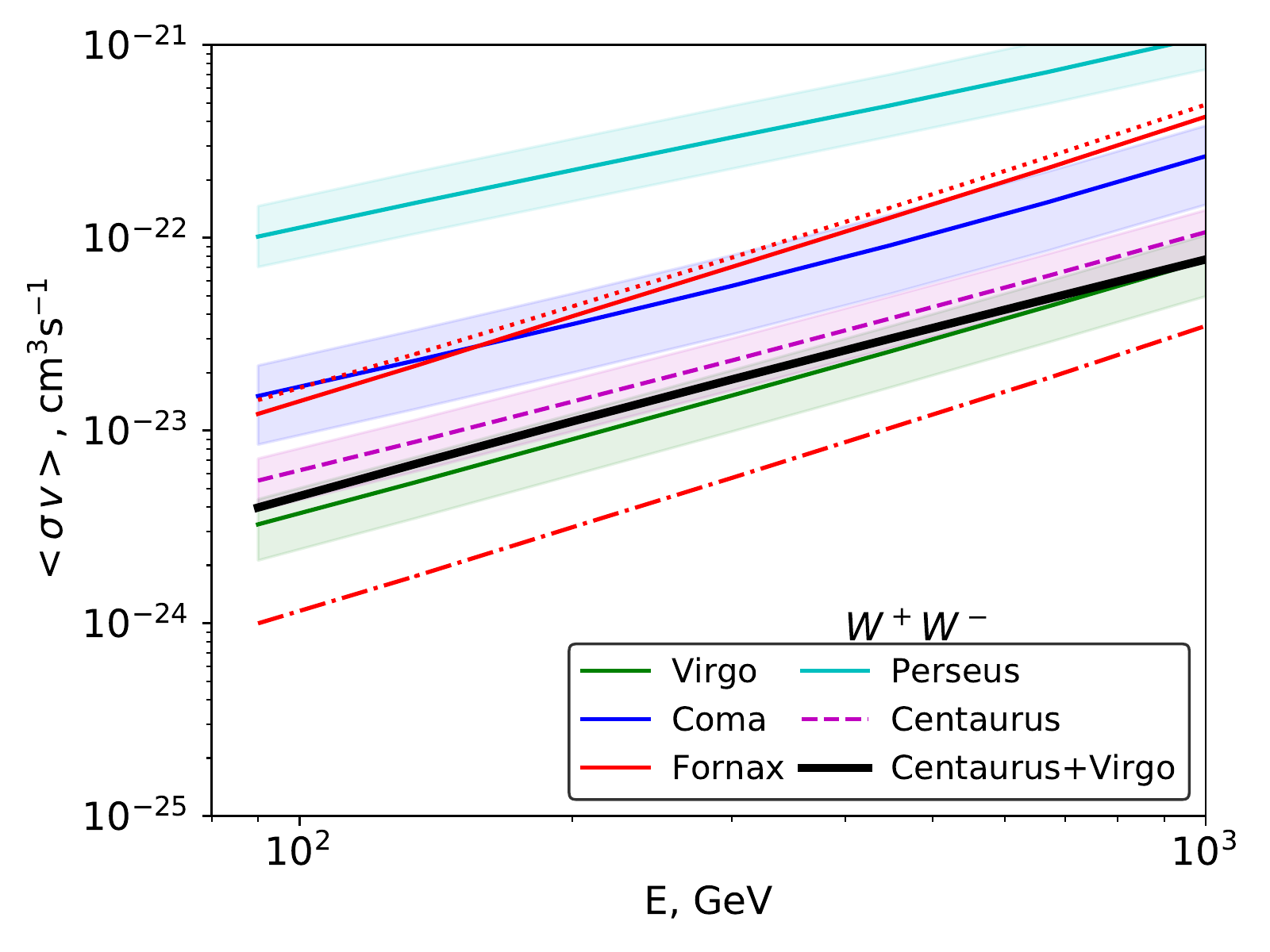}
\includegraphics[width=0.45\linewidth]{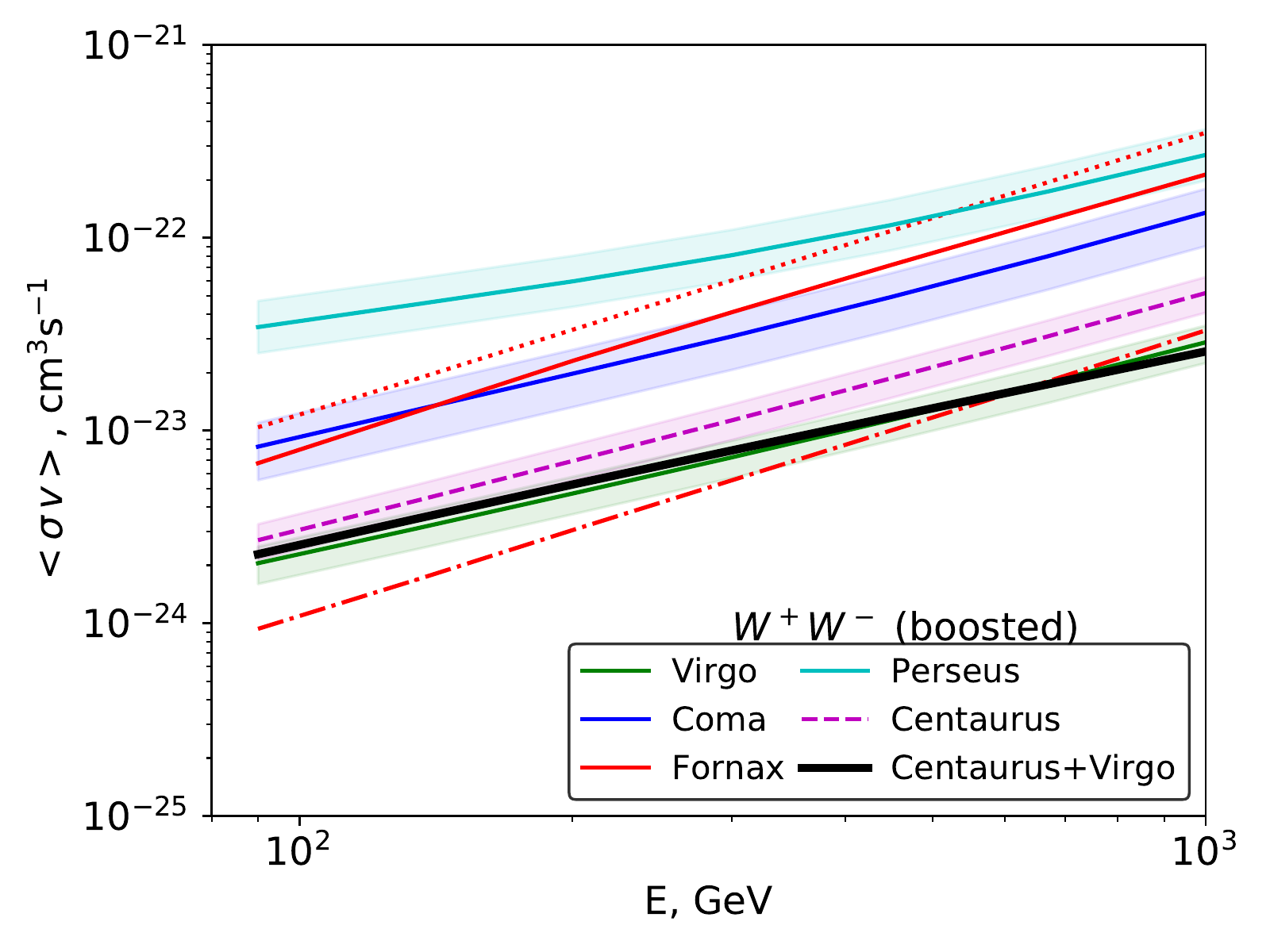}
\caption{The 95\% C.L. limits on WIMP dark matter's annihilation cross-section for the $W^+W^-$ annihilation channel, across the sample of considered galaxy clusters. See Fig.~\ref{fig:limits_bb} for detailed panel and lines description.}
\label{fig:limits_ww}
\end{figure*}
%%%%%%%%%%%%%%%%%%%%%%%%%%%%%%%%%%%%%% 
 %%%%%%%%%%%%%%%%%%%%%%%%%%%%%%%%%%%%%%
\begin{figure*}
\includegraphics[width=0.45\linewidth]{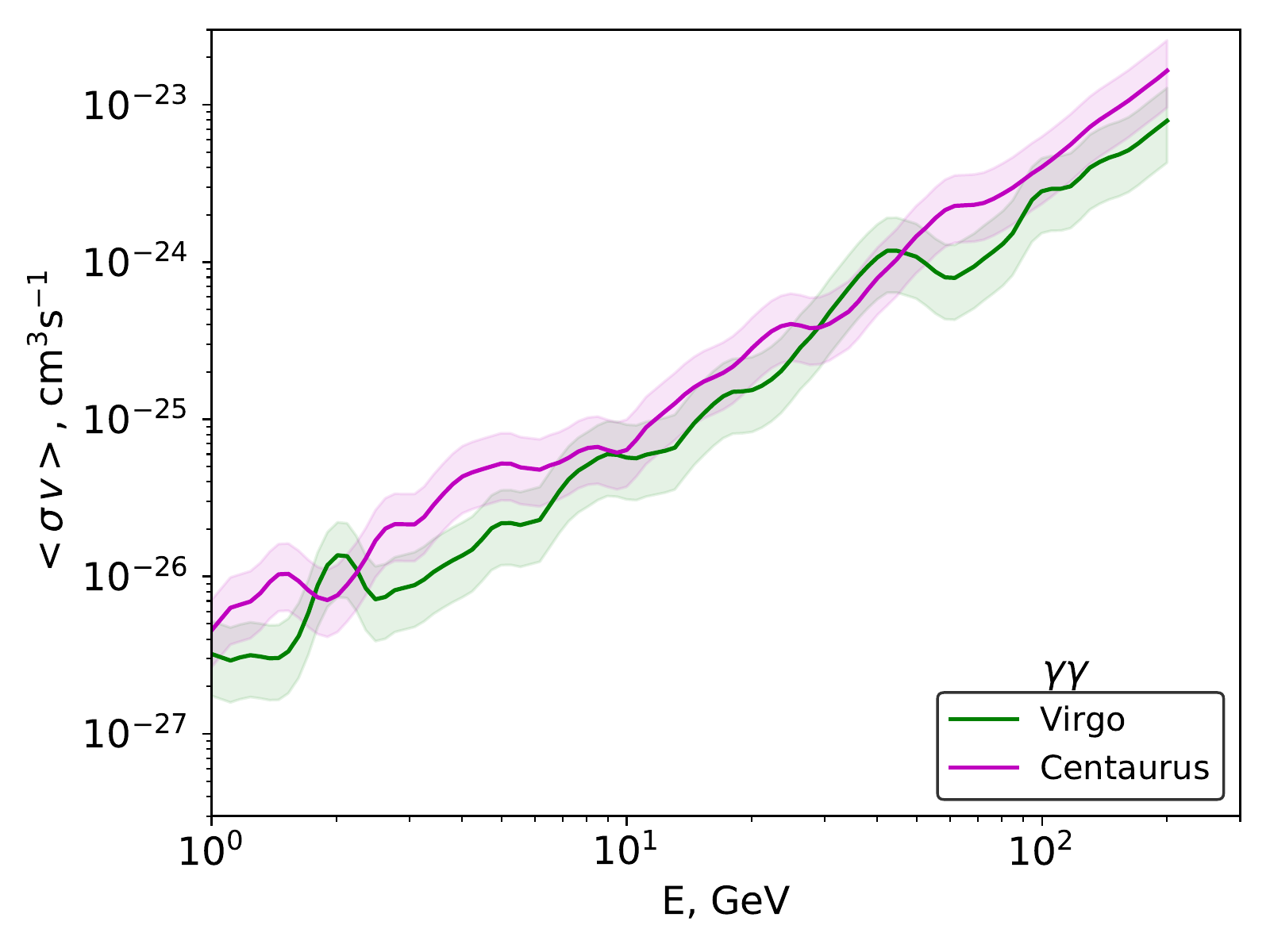}
\includegraphics[width=0.45\linewidth]{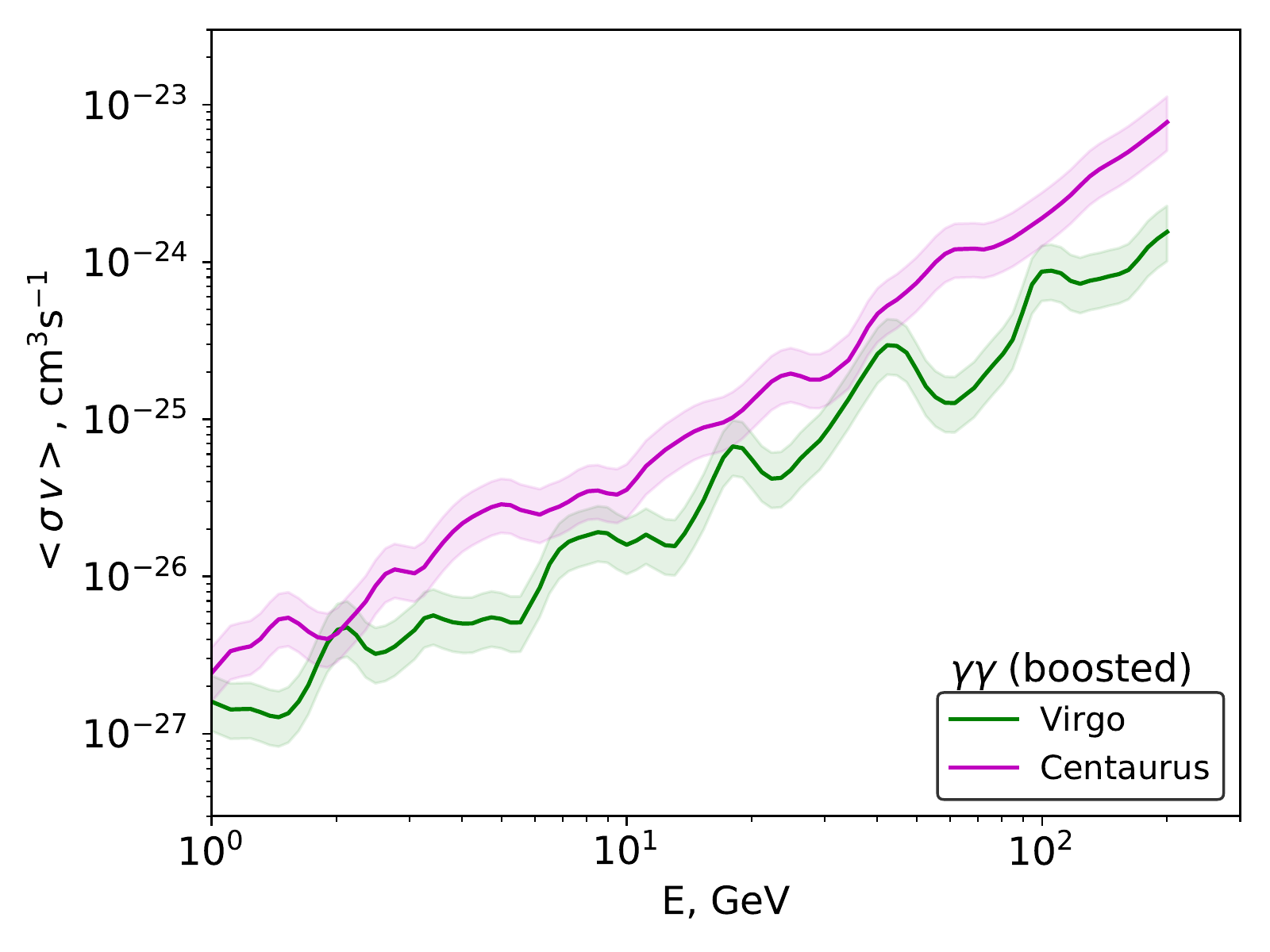}
\caption{The 95\% C.L. limits on WIMP dark matter's annihilation cross-section for the $\gamma\gamma$ annihilation channel, in the Virgo and Centaurus clusters. These clusters were selected to show unique channel, given that they provided the best limits for the $b \overline{b}$ and $W^+W^-$ channels. See Fig.~\ref{fig:limits_bb} for detailed panel and lines description.}
\label{fig:limits_gamgam}
\end{figure*}
%%%%%%%%%%%%%%%%%%%%%%%%%%%%%%%%%%%%%% 

\section{Results and Discussion}
\label{sec:results_discussion}
Galaxy clusters are the largest virialised objects in the Universe, hosting large amounts of dark matter which make them effective targets for indirect DM searches at all wavelengths. With their relatively well measured DM density profiles, low astrophysical background signal in the GeV-TeV energy range, galaxy clusters are one of the best targets for WIMP searches with current and future high-energy observational facilities~\citep{bringmann12,funk15,morselli17,yapici17}.

This study carried out an analysis, using almost 12 years of Fermi/LAT data, of five nearby clusters expected to be the strong sources of high energy photons originating from DM annihilation. We modelled the DM halo of each cluster by a spherically symmetric NFW profile with the parameters adopted from dedicated studies, see Tab.~\ref{tab:clusters}. A significant amount of attention was devoted to the accurate propagation of uncertainties on the DM profile parameters and on the uncertainties in the spatial shape and intensity of expected signal. For the Perseus and Fornax clusters, more than one measurement of DM profiles were reported in the literature; in these cases we used different available profiles to check the possible effects of systematic uncertainties on the obtained results.

This study, as well as its larger data set, also benefited from the utilisation of the updated Pass 8 Fermi/LAT instrument response functions (IRFs), characterised by better non-photon background rejection than Pass 7 IRFs used in previous studies~\citep{Xiaoyuan}.  Furthermore, as previously mentioned, this study has focused in great detail on the uncertainties in the DM profiles of the selected objects and their effect on the derived limits, where we have taken a more rigorous approach. Finally, our comparison of different profiles for the same object is a distinctive feature of this work in comparison to previous works. This comparison has allowed us to highlight potential anomalies in profiles and provide basis for further testing and comparison of these.

In absence of a significant detection ($>2.5\sigma$) of a DM annihilation signal in any of the considered clusters, we present 95\% C.L. constraints on the annihilation cross-section for the $b\overline{b}$ and $W^+W^-$ channels; channels for which the highest branching ratios are expected within a certain class of (astro)physically motivated constrained minimal supersymmetric Standard Models~\citep{Jeltema_2008}. For the two clusters showing the best constraints in these channels we, in addition, present the limits on the flux and annihilation cross-section in the $\gamma\gamma$ channel, usually considered as a ``smoking-gun'' signature of dark matter annihilation or decay in the Universe. Corresponding results are presented in Fig.~\ref{fig:limits_bb}--\ref{fig:limits_gamgam}.
Left/right panels of the figures show the results for DM annihilation cross-section in the corresponding channel, both with or without accounting for a presence of substructures in DM halo. 

Formally the tightest constrains were provided by the Fornax cluster, assuming that the dark matter distribution follows an RB02~\citep{Reiprich_2002} NFW profile (see red dot-dashed line in Figs.~\ref{fig:limits_bb}--\ref{fig:limits_ww}). Similar conclusions were previously derived by~\citet{Xiaoyuan} for the data set covering first 3 years of Fermi/LAT observations. Two other profiles reported in the literature -- DW01~\citep{dw01} and SR10A10~\citep{sr10} (solid and dotted red lines correspondingly) result in weaker constraints by a factor of $\sim 20$.
A substantial difference in the results of this profile to other considered profiles was also noticed  within HESS TeV data analysis~\citep{Abramowski_2012}. Given the large discrepancy in the limits in the Fornax cluster, we do not draw the final limits based on this object.

After the Fornax (RB02 profile), the best constraints were provided by the Virgo and Centaurus clusters. The constraints from the Coma and Perseus clusters proved to be up to an order of magnitude higher than their counterpart limits from the Fornax RB02 profile or the limits from Virgo and Centaurus. The significant degradation of limits from these clusters follows from\textit{ (i)}: somewhat lower $J$-factors in these objects ; \textit{(ii)} the presence of $\gamma$-ray sources close to the center of the cluster (NGC~1275 for the Perseus cluster and a source in the Coma cluster potentially associated with H~I cloud \ref{sec:coma}). The presence of bright sources implies partial confusion of the fluxes attributed by the analysis between the gamma ray source and DM signal, which resulted in a higher level of flux uncertainties and consequently weaker limits on the DM annihilation cross-section.

The black solid lines in Figs~\ref{fig:limits_bb}--\ref{fig:limits_ww} show the best results from  stacked analyses of the Centaurus and Virgo clusters. To perform such stacking, we built the log-likelihood profile as a function of $\langle\sigma v\rangle$ for each of these clusters. The limits shown correspond to the $\langle\sigma v\rangle$ value at which the sum of log-likelihood profiles changes by $2.71/2$~\citep{rolke05}\footnote{See also \href{https://fermi.gsfc.nasa.gov/ssc/data/analysis/scitools/upper\_limits.html}{discussion on Fermi/LAT upper limits calculation}. }.

Dashed black lines in the left panels of Figs~\ref{fig:limits_bb}--\ref{fig:limits_ww} correspond to the~\citet{Xiaoyuan} limits (without accounting for substructures), for each of the channels. %We note, that the corresponding~\citet{Xiaoyuan} limits accounting for the presence of substructures are by about 2 orders of magnitude better than their analogues without substructures. In our results the improvement is much more modest (a factor of a few) which is connected to more conservative estimation of the boost factor in this work ($\sim 10$ vs. $\gtrsim 10^3$ considered in~\citet{Xiaoyuan}).

Despite utilising substantially larger amounts of data, the obtained limits are comparable to ones obtained by~\citep{Xiaoyuan} for smooth DM profiles and weaker by a few orders of magnitude for profiles accounting for the presence of substructures. We argue that such discrepancies arise from two main differences in the methods of analysis chosen by this study and~\citet{Xiaoyuan}. The best limits within~\citet{Xiaoyuan} originate from the Fornax cluster. The dark matter profiles presented in the literature for this cluster (see Tab.~\ref{tab:clusters}) result in a difference of one order of magnitude on limits on DM annihilation cross-sections, see red solid, dashed and dot-dashed Fig.~\ref{fig:limits_bb}-\ref{fig:limits_ww} corresponding to the DW01, SR10A10 and RB02 profiles. The results presented in~\citet{Xiaoyuan} are based on the RB02 profile (dot-dashed curve) which formally provides the best limits in this work too. Given the large discrepancy between  the results of differing profiles for the Fornax cluster, in contrast to ~\citet{Xiaoyuan}, we have opted not to draw any conclusions based on this cluster. For the profiles characterised by a presence of subhalos we assumed conservative values of boost-factors (total $J$-factor boost $\sim 10$ vs $\sim 1000$ considered by~\citet{Xiaoyuan}). The combination of these two factors resulted in the correspondingly weaker upper limits.

Although the presented constraints are somewhat weaker, they are within an order of magnitude comparable to the recent limits based on a stacked analysis of a large number of dwarf spheroidal galaxies~\citep{fermi_dsphs15,hoof20}. However, the potential uncertainty connected to stacking  large numbers of dSphs with poorly known individual profiles can reach up to a factor of 2~\citep{linden20}, which has the potential to make the presented results more competitive.
We note, that systematic uncertainties in DM density profiles of clusters of galaxies can affect this study to a comparable extent, see e.g.~Figs.~\ref{fig:limits_bb}--\ref{fig:limits_gamgam}. At the same time we argue that the results derived in this work  are based on a substantially different class of objects containing DM than dSphs. This make the results all the more important as they provide grounds for cross-check analyses.

We finally note the importance of accurate measurements of DM profiles in galaxy clusters. The accurate measurement of such profiles (including explicit quotation of correlation between the parameters) with currently available X-ray and/or optical data is key for the reliability in the estimation of $J$-factors, and consequently, the parameters of annihilating dark matter.

\noindent\section*{Acknowledgements}
The authors acknowledge support by the state of Baden-W\"urttemberg through bwHPC. This work was supported by DFG through the grant MA 7807/2-1.

\noindent\section*{Data availability}
The data underlying this article are available at the Fermi LAT data server, https://fermi.gsfc.nasa.gov/cgi-bin/ssc/LAT/LATDataQuery.cgi. Other data used in the article will be shared on reasonable request to the corresponding author.

\bibliographystyle{mnras}
\bibliography{biblio.bib}

% Don't change these lines
\bsp	% typesetting comment
\label{lastpage}
\end{document}